# Optical Properties of Indium-Gallium-Oxide Microcrystalline Alloy Films: From the Visible to the Deep-UV

HM Borhanul Alam[1], Dipak Oli[1], You Qiang[1], Bisheswor Acharya[1], Jesse Huso[2], Matthew D. McCluskey[3], and Leah Bergman[1]

[1]Department of Physics, University of Idaho, Moscow ID 83844 USA

[2]Klar Scientific, Pullman WA 99163 USA

[3]Department of Physics and Astronomy, Washington State University, Pullman WA 99164 USA

Corresponding author: Leah Bergman, Lbergman@uidaho.edu

## Abstract

The tailored optical properties of $(In_xGa_{1-x})_2O_3$ microcrystalline films were studied as a function of composition x via transmission, Urbach energy analysis, and spatial photoluminescence (PL) mapping of the self-trapped hole (STH) emission, with the objective of addressing material characteristics specific to this alloy system. Up to x = 0.46, the optical gap exhibited a redshift of 1 eV from the deep to the near-UV range, while the STH PL was redshifted by 0.5 eV in the visible range. For higher composition, x = 0.63, the transmission spectra indicated the co-existence of two optical gaps attributed to Ga-rich and to In-rich domains, implying that this sample is phase-separated. However, the saturation behavior of the optical gap and that of the STH PL showed that incipient phase separation occurs at a lower composition: x ~ 0.3. This is consistent with the compositional trend found for Urbach energy, implying that phase segregation in the alloys is a major defect even at its incipient stages. Additionally, Urbach analysis of $(In_xGa_{1-x})_2O_3$ was compared to that of $Mg_xZn_{1-x}O$. Both systems were found to have similar compositional dependence: at lower range, Urbach energies exhibited a negligible increase, while at the higher range a significant dependence on the composition was found. The main difference between the two alloy systems is in their Urbach energy: those for $(In_xGa_{1-x})_2O_3$ were significantly larger than those of $Mg_xZn_{1-x}O$. This stems from the strong hole coupling to phonons of $(In_xGa_{1-x})_2O_3$, which provides a dynamic transition additionally to that of defect-type.

# 1. Introduction

$\beta$-Ga$_2$O$_3$ has bandgap in the deep-UV range of ~ 5 eV, while that of In$_2$O$_3$ is in the visible range ~ 2.7 eV [1-2]. These two semiconductors have been established to have many viable applications in UV optoelectronic technologies [3-4]. To extend their range of application, creating the alloy system (In$_x$Ga$_{1-x}$)$_2$O$_3$ may be advantageous for the realization of by-design light sources and sensors, as well as for heterostructure and bandgap engineered devices from the visible to deep-UV [5-7]. In the alloy formula, x is the alloy composition and ranges over $0 \leq x \leq 1$: varying this parameter is what results in the tailored properties. Due to the different crystal structures of the two end members, monoclinic for $\beta$-Ga$_2$O$_3$ and cubic bixbyite for In$_2$O$_3$, total solubility across the entire composition range cannot be achieved, and at certain composition the alloy is expected to no longer be a single phase and should exhibit phase separation in a sample where the two structures co-exist [8]. The latter modifies extensively the optical characteristic of such samples.

The photoluminescence (PL) of (In$_x$Ga$_{1-x}$)$_2$O$_3$ is an interesting issue because $\beta$-Ga$_2$O$_3$ cannot have bandgap deep-UV emission ~ 5 eV due to self-trapping of the holes at the oxygen site, a consequence of the strong hole-phonon coupling of this material. Usually, the main efficient light emission in $\beta$-Ga$_2$O$_3$ is that of the self-trapped hole (STH) at ~ 3 eV [2, 9]. Moreover, theoretical studies have indicated that the light emission of STH in In$_2$O$_3$ is expected to be ~ 1.5 eV [2], thus alloying may enable tunable luminescence from the near-IR to the near-UV. Specifically, micro crystalline (In$_x$Ga$_{1-x}$)$_2$O$_3$ films and microstructures have been investigated previously for applications such as UV-radiation detectors and gas sensors, that due to their large surface to volume ratio can be rendered to be highly efficient [5, 7, 10 - 12]. Thus, for heterostructure and bandgap engineered technologies, understanding in-depth the basic properties of this alloy system in its polycrystalline form is of merit.

In this work we present optical studies of microcrystalline (In$_x$Ga$_{1-x}$)$_2$O$_3$ films for composition x = 0, 0.075, 0.19, 0.31, 0.39. 0.46, 0.63, and 1. Up to x = 0.46 the optical gap redshifted by more than 1 eV, and at x = 0.63 the film was found to be phase separated with two distinct optical gaps corresponding to Ga-rich and In-rich phases, respectively. The nature of the optical gaps was analyzed in terms of their Urbach energies, which were found to be extremely large relative to the previously studied magnesium zinc oxide (Mg$_x$Zn$_{1-x}$O) alloy system. This finding is discussed in terms of the strong hole–phonon coupling of $(In_xGa_{1-x})_2O_3$ that contributes a component to Urbach energy in addition to that of defects. Moreover, the dependence of optical gap and that of the STH PL on the alloy composition was found to obey a compositional saturation type behavior for which incipient phase separation was inferred to take place at composition x~ 0.3. At this composition, Urbach energy exhibited an increase, implying that phase separation is a major defect in the alloys.

## 2. Experimental Approach

The $(In_xGa_{1-x})_2O_3$ alloy films were synthesized on quartz substrates via a wet chemical approach. Indium (III) nitrate hydrate and gallium (III) nitrate hydrate powders, with purity 99.99%, purchased from Thermo Scientific, U.S.A., were used as the precursor chemicals. Solutions were created using deionized water and were dispersed over the quartz substrate, which was then heated on a hot plate at 190 C for 30 minutes, followed by annealing at 1100 C for 1 hour in the ambient environment. An annealing was performed using a Lindberg/Blue M quartz tube furnace controlled by a Yokogawa UP-150 temperature controller. The alloy compositions of the films were determined by energy dispersive X-ray spectroscopy (EDS) using a Zeiss Supra 35 scanning electron microscope. The $(In_xGa_{1-x})_2O_3$ alloy films have compositions of x = 0, 0.075, 0.19, 0.31, 0.39, 0.46, 0.63, and 1.

Ultraviolet-Visible (UV-Vis) transmission measurements were performed at room temperature employing an Agilent Cary 300 double-beam spectrophotometer in the wavelength range of 190 – 700 nm (6.5 – 1.8 eV). The data were analyzed using Origin software. Photoluminescence spectra and images of the $(In_xGa_{1-x})_2O_3$ alloy films with x = 0, 0.19, and 0.39, were acquired and analyzed using a Klar Scientific Mini Pro microscope with a 266 nm ( 4.7 eV) laser, a detected spectral range of 200 – 648 nm (6.2 - 1.9 eV), a laser spot size of 650 nm, and spatial step size of 10 μm. The surface morphologies of two representative samples with composition x = 0.19 and 0.39 were investigated using a TT-AFM atomic force microscope.

## 3. Results and Discussion

The transmission spectra of the $(In_xGa_{1-x})_2O_3$ for several alloy compositions is presented in Figure 1 showing the redshift trend as a function of increasing indium composition, while Figure 2 presents the optical gap of the alloys. The optical gaps of the films were ascertained via the derivative of the transmission method that was proven previously to be useful for analyzing spectra with large in-gap tails, and moreover it agreed with the Tauc method [13]. The optical gap of the $Ga_2O_3$ is found to be ~ 4.86 eV, and that of the alloy with indium composition x = 0.46 is at 3.78 eV. For that group of alloys an optical gap shift of ~ 1.08 eV was achieved. For similar composition range, a comparable optical shift was observed for the alloy system grown via a pulsed laser deposition (PLD) as well as a sol-gel method [14 -15].

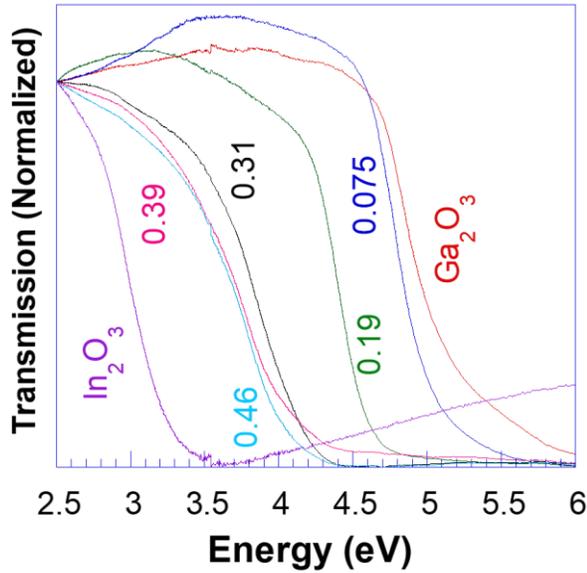
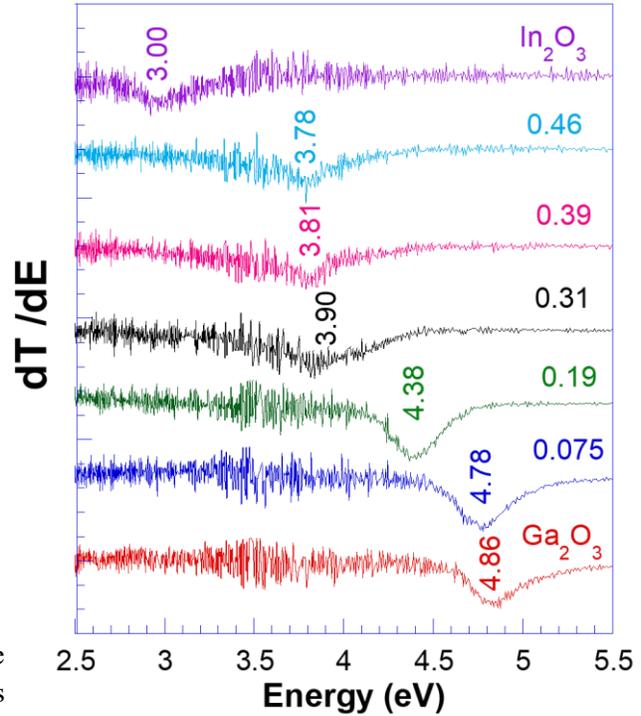

**Fig 1.** The normalized transmission spectra for the two end members of the alloy: $Ga_2O_3$, and $In_2O_3$, as well as for the alloys of composition: x = 0.075, 0.19, 0.31, 0.39, and 0.46.

**Fig 2.** The optical gaps of the samples obtained from the derivative of the transmission spectra.

Figure 3 summarizes the optical gap behavior as a function of alloy composition. Also included in Figure 3 are the characteristics of a film with high indium composition x = 0.63. As can be seen in the inset of Figure 3, this film exhibits two signatures of optical gaps: one at ~ 4.14 eV and a weaker one ~ 3.1 eV, indicating that this alloy is phase separated into Ga-rich and In-rich domains, respectively. According to theoretical predictions, the $(In_xGa_{1-x})_2O_3$ alloy system exhibits a solubility limit at x ~ 0.5 which is due to the two different crystal structures of the alloy's end members [1]. The Ga atoms in the monoclinic $Ga_2O_3$ structure occupy the octahedral as well as the tetrahedral sites, while the indium atoms in the bixbyte $In_2O_3$ structure have only the octahedral site symmetry. According to reference [1] the solubility limit stems from energetically favorable site symmetry for the alloy constituents. Specifically, for x ~ 0.5 or less, most indium atoms are in the favorable octahedral sites of the $(In_xGa_{1-x})_2O_3$ alloy, while above that concentration, the indium atoms reside in the tetrahedral sites which are energetically unfavorable [1]. Experimental studies on the phase equilibria in $Ga_2O_3$-$In_2O_3$ ceramics, at the temperature range of 1000 C -1400 C which is similar to our growth temperature, have found the solubility limit to be x ~ 0.45 [8], consistent with the theoretical work of reference [1].

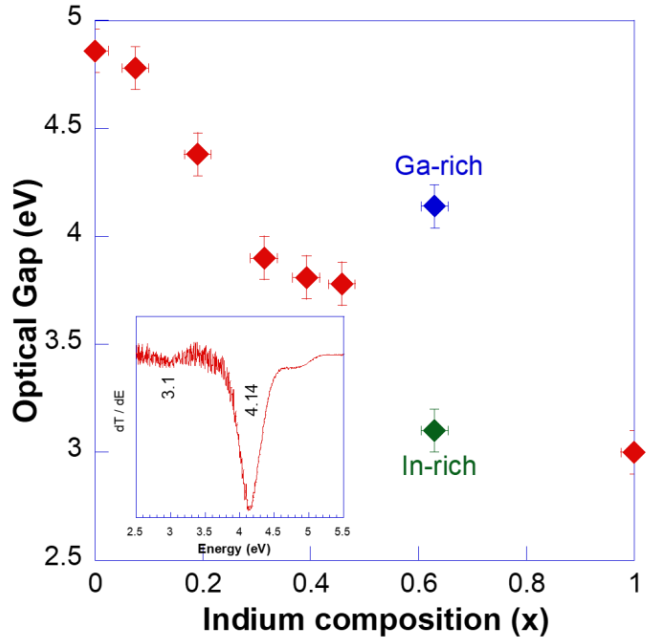 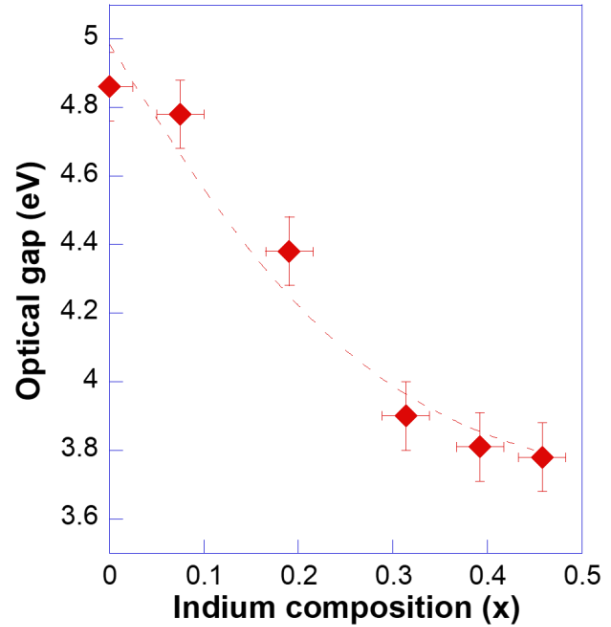

**Fig 3.** The behavior of the optical gap as a function of composition. Also included in the inset is the optical characteristic of a sample with high indium composition x = 0.63 that exhibits two optical gaps weak one at 3.1 eV attributed to In-rich domains and one at the UV range of 4.14 eV due to Ga-rich domains.

**Fig 4.** The optical gap as a function of composition (diamonds) and a fit of the data using the saturation model of Equation 1 (line) for which a saturation composition x ~ 0.3 was ascertained.

Studies of $(In_xGa_{1-x})_2O_3$ grown via PLD found that the onset to phase separation takes place at x ~ 0.3 where the structures of the two end members co-exist as well as an additional hexagonal structure [16]. Moreover, that study found that the hexagonal direct optical gap is at 4.13 eV, and somewhat lower for the indirect gap, which is similar to the optical gap found here for x = 0.63 as is presented in Figure 3. In other studies regarding $(In_xGa_{1-x})_2O_3$ films grown via molecular beam epitaxy, it was found that at a growth temperature of ~ 800 C the solubility limit of indium was marginal at ~ 0.08, but for lower growth temperature ~ 600 C the limit was significantly raised to x = 0.35 [17].

The optical gap trend can be utilized to investigate the composition corresponding to the onset of the phase separation. For this, the saturation model of the following form was employed [18 - 19]:

$$E_{OG} = E_0 - A \tanh\left(\frac{x}{x_0}\right). \qquad (1)$$

In Equation 1 $E_0$ is the optical gap of $Ga_2O_3$, $A$ is a constant, and $x_0$ is the saturation value that indicated the onset composition.

Figure 4 presents a fit to the data using Equation 1, for which a saturation composition of $x_0 \sim 0.3$ was obtained that can be interpreted as the onset of the solubility limit for our films. As can be seen in Figure 4, up to $x \sim 0.3$ the optical gap exhibits a monotonic linear decrease; above that value there is a weaker response of the optical gap to the increasing indium composition indicating the gradual inability of the monoclinic lattice to accommodate the indium atoms. Similar behavior was previously observed in $Mg_xZn_{1-x}O$ thin film for which the solubility limit was analyzed via the saturation trend of the LO-Raman mode frequency [19]. As is discussed below, due to the non-allowed transition of $In_2O_3$ it is challenging to observe the optical gap of the In-rich precipitates. However, at sufficient indium content such as $x = 0.63$ (see the inset to Figure 3), a weak signature is detected in the spectrum.

Although not the focus of this paper, it bears noting the optical characteristic of the $In_2O_3$ film. The fundamental bandgap of $In_2O_3$ was predicted to be at the $\sim 2.6 - 2.9$ eV range, and its optical transitions at the $\Gamma$ point are forbidden [20 - 21]. Previous research into that topic has observed an onset of optical transitions at the range of $\sim 3 - 3.7$ eV, and were attributed to allowed transitions that involve allowed lower valence bands at $\Gamma$ point [20]. To gain further insight into the behavior of the optical gap of the $In_2O_3$ film, we employed transmission spectroscopy that encompassed a lower energy range, as is presented in Figure 5. It can be seen in Figure 5 that two absorption events are taking place: one at $\sim 3$ eV and a weaker one at $\sim 2.3$ eV. We attribute the 3 eV spectral feature to the allowed transition, while the weak absorption at $\sim 2.3$ eV is attributed to the optical gap of the forbidden transitions. We suggest that due to the inherent structural defects of the granular film the selection rules were relaxed, resulting in an observable weak transmission. Also, our values of optical gap are lower in energy than the actual bandgaps of this alloy system. This is due to transitions of in-gap states arising from factors such as structural defects and alloy inhomogeneities, as well as to coupling of free carriers to phonons which is significant in

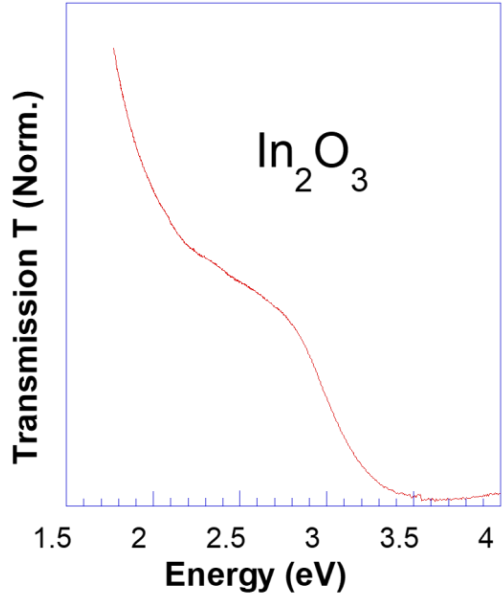
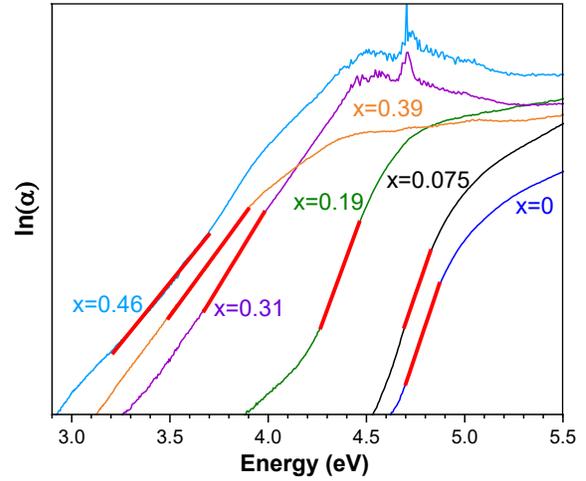

**Fig 5.** The transmission spectrum of the In$_2$O$_3$ sample that includes the visible range.

**Fig 6.** Urbach model plots of the samples of composition x = 0, 0.075, 0.19, 0.39, and 0.46. For each plot, the straight line below the optical gap is a fit to Equation 2 from which the Urbach energy was ascertained.

(In$_x$Ga$_{1-x}$)$_2$O$_3$. This assertion will be discussed in the following sections within the frame of the Urbach model as well as by examining the optical characteristics of the STH in our films.

As can be seen in Figure 6, the absorption spectra, α, of the samples exhibit a band tail which can be analyzed via the Urbach model expressed in Equation 2 [9, 22]

$$\alpha(E) = \alpha_o e^{\frac{E}{E_u}} \qquad (2)$$

In Figure 6, ln (α) for the films with indium composition up to the phase separation region, i.e., x = 0 to x = 0.46, is presented. From the fit to the linear region around the energy values of the optical gap, Urbach energies, $E_u$, were ascertained and are presented in Figure 7. In Figure 7, for comparison purposes Urbach energies for the Mg$_x$Zn$_{1-x}$O alloy system that was grown previously via a sputtering technique and has nanocrystalline morphology are also presented [23]. The trend in Urbach energy as a function of alloying for both alloy systems is quite similar: at the low regime of alloying, it is almost constant, while at higher values of alloying the change in Urbach energies becomes significant. These findings indicate that the impact of alloying takes place at relatively high compositions. At these compositions more structural defects are expected to be present due to alloy inhomogeneities and incipient phase separation, the latter which will be discussed in a later section. To get an insight to the morphological changes upon alloying, atomic force microscopy (AFM) was utilized, and the images are presented in Figure 8. Two representative regimes of composition were chosen: one at x = 0.19 the another at x = 0.39. As can be seen in Figure 8, the morphology of the low compositions film consists of relatively sharp micro-structures, while at the high composition the morphology of the film is rather smooth which is a characteristic of a disordered or amorphous-like material. As was discussed in [9], the granular morphology contributes an underlying defect component to the Urbach energy.

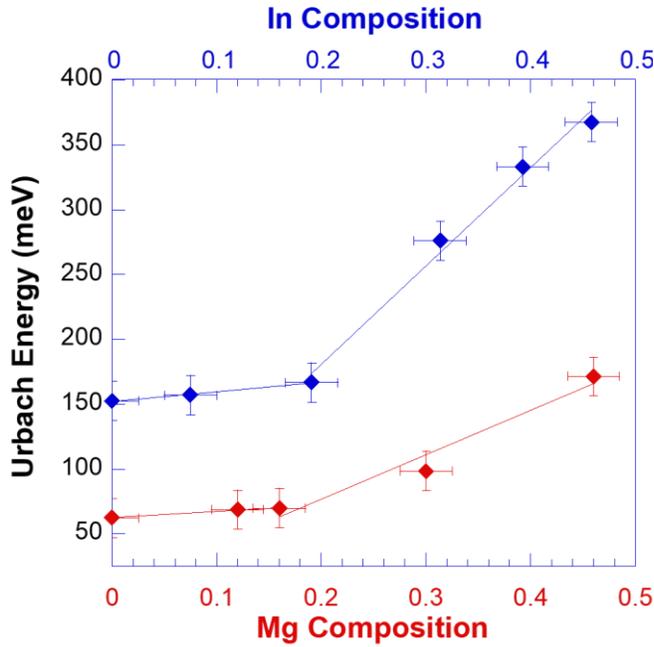
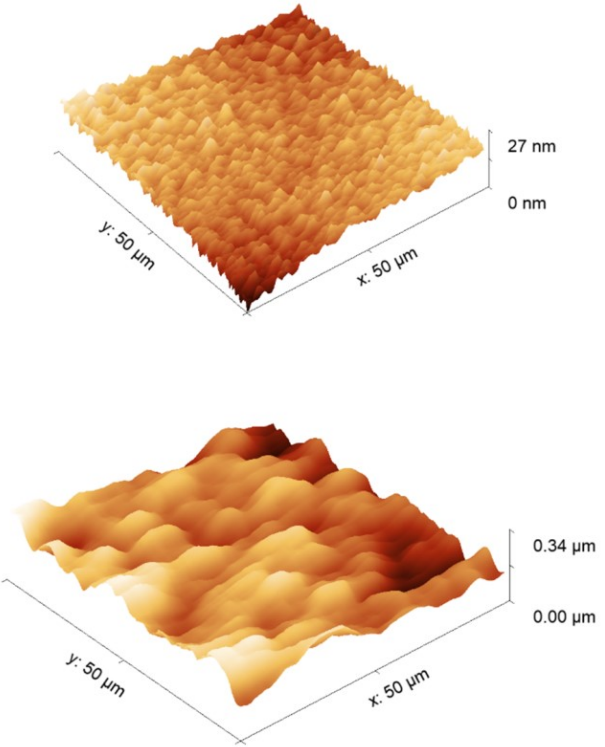

**Fig 7.** Urbach energies as a function for composition for two alloy systems; $Mg_xZn_{1-x}O$ (red) and $(In_xGa_{1-x})_2O_3$ (blue). The films of these two alloy systems exhibit similar granular morphology.

**Fig 8.** AFM images for x = 0.19 (top) and x = 0.39 (bottom), showing the granular morphology of the films.

A striking difference between the two alloy systems is the difference in the values of the Urbach energies, which are significantly larger for $(In_xGa_{1-x})_2O_3$ as can be observed in Figure 7. Large Urbach energies up to 150 meV of β-$Ga_2O_3$ for films and crystals were previously reported [24]. The researchers discussed these large values in terms of the strong hole-phonon coupling and the self-trapping of charge carriers [24]. This dynamic for the case of a strong self-trapping was theoretically considered in reference [25] and was attributed to optical transitions of carriers to the many vibrational states, which should result in low energy tails below the bandgap [25]. β-$Ga_2O_3$ has extremely strong hole-phonon coupling, somewhat less for $In_2O_3$ and negligible for MgO, while ZnO has none [2]. Thus, in addition to the impact of defects, we attribute the extremely large values of Urbach energies in our $(In_xGa_{1-x})_2O_3$ alloys to originate from its strong hole-phonon dynamics. In the following section, a study of the self-tapping characteristics of $(In_xGa_{1-x})_2O_3$ is presented, as was studied via the luminescence of the STH.

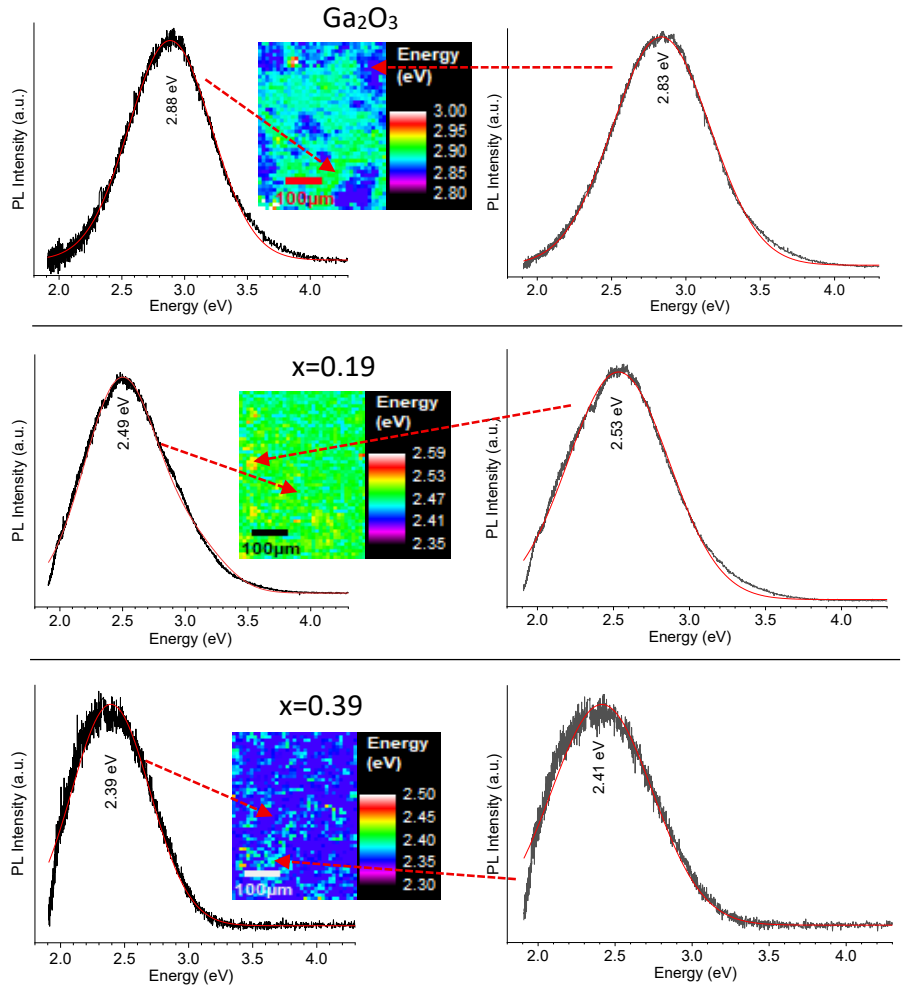

**Fig 9.** The scanning PL of the self-trapped hole for $Ga_2O_3$ and for Indium composition for x = 0.19 and x = 0.39. For each sample, the spectra represent the two dominant energy regions. The spectral lineshape is fitted by a Gaussian function.

One of the most interesting crystal dynamics properties of $Ga_2O_3$ is its self-trapped hole (STH) that inhibits bandgap luminescence at the deep UV range ~ 5 eV. In their theoretical study, Varley et al. predicted that due to the large hole-phonon coupling of this alloy, the holes become highly localized [2]. This self-trapping impedes the deep UV photoluminescence of the free e-h pair recombination at the bandgap energy range. In contrast, when a free electron from the conduction band, or an exciton, recombines with STH, a characteristic light emission at the near UV ~ 3.1 to 3.5 eV is expected which can be very dominant even at room temperature [9, 26].

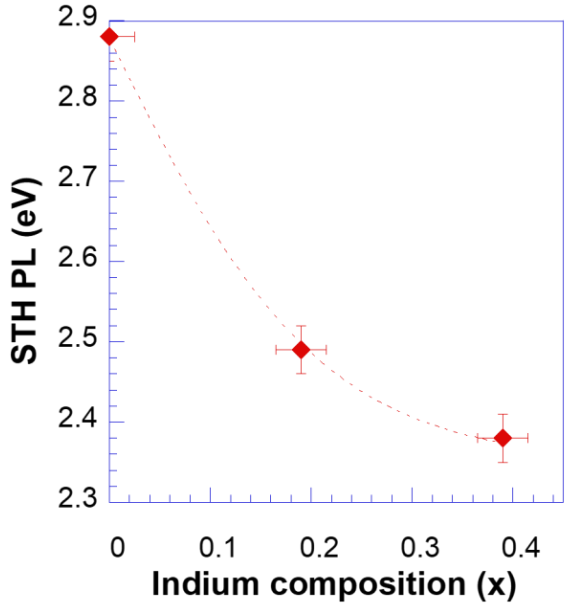

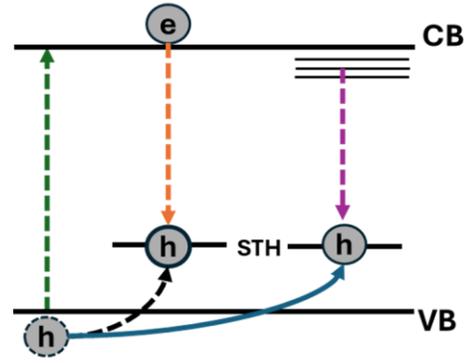

**Fig 11.** The band-energy diagram representing the recombination process of the electron with the self-trapped hole, without defect states (left) and with defect states (right, as inferred from this study).

**Fig 10.** PL peak energy of the self-trapped hole as a function of indium composition. The line is a fit of the saturation model in Equation 1.

Figure 9 presents a spatial scanning of STH luminescence for films with indium composition x = 0, 0.19, and 0.39, while in Figure 10 the dominant STH PL versus composition is plotted. The STH of $Ga_2O_3$ was found to have an average energy position ~ 2.9 eV, that with x = 0.19 is at 2.49 eV, and is at 2.39 eV for x = 0.39. An immediate conclusion is that as the indium composition increases the STH PL energy shifts to lower energies. This is an expected behavior since the theoretical value of the STH in $In_2O_3$ was reported previously to be at 1.51 eV [2].

Moreover, as can be seen in Figure 10, the STH PL behavior as a function of composition is similar to that of the optical gap presented in Figure 4. To check for consistency, a fit to the saturation model of Equation 1 was performed and is presented in Figure 10. The fit yielded a saturation composition x ~ 0.21 which up to experimental approach is similar to that of the optical gap at x ~ 0.3. This indicates that an incipient phase separation in the films studied here takes place in the approximate composition of x = 0.2 – 0.3. At a similar composition, Urbach energies of the $(In_xGa_{1-x})_2O_3$ alloy system were found to undergo a change in their compositional dependence (see Figure 7). Up to that composition, a weak dependence was found, while above that composition Urbach energies exhibited a strong response to alloying. The above results imply that phase separation in the alloy matrix constituents is of a detrimental defect-type.

The STH of our $Ga_2O_3$ film is at ~ 2.88 eV, which is a lower energy than the expected values previously reported to be ~ 3.1 eV to 3.5 eV [2, 9, 27 - 30]. We attribute the lower energy of STH to originating from electron transitions from defect states below the conduction band to the STH energy level. Figure11 schematically presents the STH photoluminescence dynamics of our $Ga_2O_3$ film. As pertains to film quality, as can be seen in the images of Figure 9, there is no significant spatial fluctuation of the STH PL energy across a sample, indicating the uniformity of the alloying process.

## 4. Conclusions

$(In_xGa_{1-x})_2O_3$ microcrystalline thin films with tailored optical properties spanning the visible to the deep-UV range were realized. The light emission of the STH and the optical gaps were found to be redshifted in a nonlinear fashion with indium composition. This trend was addressed via the saturation model which indicated that the onset to phase separation occurs at relatively low composition x ~ 0.3; above that, the optical gap and the STH PL were found to have weak response to the indium. Furthermore, at a similar composition, Urbach energies of the $(In_xGa_{1-x})_2O_3$ alloy system were found to drastically change their compositional dependence. Specifically, up to that point, a negligible dependence was found, while above that composition Urbach energies exhibited a strong response to alloying, implying that phase segregated precipitates are a major defect-type even at initial stages. Moreover, compared to another ternary alloy, $Mg_xZn_{1-x}O$, the Urbach energies of $(In_xGa_{1-x})_2O_3,$ were significantly larger. This was discussed in terms of the strong hole-phonon coupling characteristics of this alloy system, which in addition to the impact of defects, broadens the Urbach tail.

## Acknowledgements

This research was supported by the U.S. Department of Energy, Office of Basic Energy Sciences, Division of Materials Science and Engineering under Award No. DE-FG02-07ER46386.

## Author Declarations

The authors have no conflicts to disclose.

## Data Availability Statement

The data that supports the findings of this study are available within the article.